\documentclass{IEEEtran4PSCC}
\ifCLASSINFOpdf
   \usepackage[pdftex]{graphicx}
\else
   \usepackage[dvips]{graphicx}
\fi
\usepackage[cmex10]{amsmath}

\usepackage{eurosym}
\usepackage{graphicx}
\usepackage{booktabs}
\usepackage{algorithm}
\usepackage{algpseudocode}
\usepackage{amsmath}
\usepackage{amssymb}

\usepackage[T1]{fontenc}
\usepackage[utf8]{inputenc}
\usepackage{csquotes}

\usepackage[backend=biber,style=ieee,sorting=none,sortcites=true]{biblatex}

\usepackage[normalem]{ulem}
\usepackage{xcolor}
\usepackage[table]{xcolor}
\definecolor{darkgreen}{RGB}{0, 128, 0}
\definecolor{darkred}{RGB}{139, 0, 0}
\definecolor{delftblue}{RGB}{0,166,214}
\definecolor{custombeige}{HTML}{D0CCBE}

\usepackage{hyperref}

\makeatletter
\let\old@ps@headings\ps@headings
\let\old@ps@IEEEtitlepagestyle\ps@IEEEtitlepagestyle
\def\psccfooter#1{%
    \def\ps@headings{%
        \old@ps@headings%
        \def\@oddfoot{\strut\hfill#1\hfill\strut}%
        \def\@evenfoot{\strut\hfill#1\hfill\strut}%
    }%
    \def\ps@IEEEtitlepagestyle{%
        \old@ps@IEEEtitlepagestyle%
        \def\@oddfoot{\strut\hfill#1\hfill\strut}%
        \def\@evenfoot{\strut\hfill#1\hfill\strut}%
    }%
    \ps@headings%
}
\makeatother

\psccfooter{%
        \parbox{\textwidth}{\hrulefill \\ \small{24th Power Systems Computation Conference} \hfill \begin{minipage}{0.2\textwidth}\centering \vspace*{4pt} \includegraphics[scale=0.06]{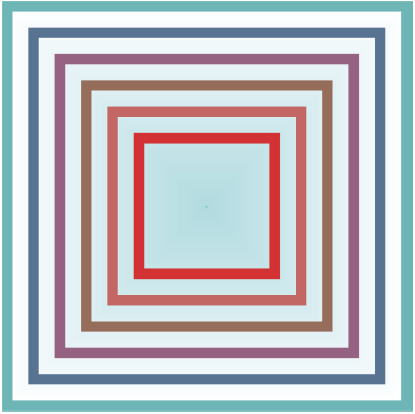}\\\small{PSCC 2026} \end{minipage} \hfill \small{Limassol, Cyprus --- June 8-12, 2026}}%
}

\begin{document}

\title{Orderbook Feature Learning and Asymmetric Generalization in Intraday Electricity Markets
}

\author{
\IEEEauthorblockN{
Runyao Yu\IEEEauthorrefmark{1}\IEEEauthorrefmark{2}, 
Ruochen Wu\IEEEauthorrefmark{1}, 
Yongsheng Han\IEEEauthorrefmark{1}, 
Jochen L. Cremer\IEEEauthorrefmark{1}\IEEEauthorrefmark{2}
}
\IEEEauthorblockA{\IEEEauthorrefmark{1} Delft University of Technology, Delft, The Netherlands}
\IEEEauthorblockA{\IEEEauthorrefmark{2} Austrian Institute of Technology, Vienna, Austria}
}

\maketitle

\begin{abstract}
Accurate probabilistic forecasting of intraday electricity prices is critical for market participants to inform trading decisions. Existing studies rely on specific domain features, such as Volume-Weighted Average Price (VWAP) and the last price. However, the rich information in the orderbook remains underexplored. Furthermore, these approaches are often developed within a single country and product type, making it unclear whether the approaches are generalizable. In this paper, we extract 384 features from the orderbook and identify a set of powerful features via feature selection. Based on selected features, we present a comprehensive benchmark using classical statistical models, tree-based ensembles, and deep learning models across two countries (Germany and Austria) and two product types (60-min and 15-min). We further perform a systematic generalization study across countries and product types, from which we reveal an asymmetric generalization phenomenon:  models trained on more liquid markets or products transfer well to less liquid ones, whereas the reverse transfer leads to substantial performance degradation.
The project page is at \textcolor{orange}{\url{https://runyao-yu.github.io/AsymGen/}}.
\end{abstract}


\section{Introduction}
\label{introduction}

Accurate probabilistic forecasting of intraday electricity price plays a vital role in enhancing decision-making for market participants under uncertainties \cite{Ensemble}. In continuous intraday (CID) markets, several studies have identified the volume-weighted average price (VWAP) from the most recent 15 minutes as a strong predictor of the future price index (ID$_3$), \textcolor{black}{an exchange-defined reference price computed as the VWAP of all matched trades in the final three hours before delivery}~\cite{beating, Econometric, Trading}. 
Previous works even argue that the last price already reflects past information, assuming weak-form efficiency~\cite{fama1970efficient}, and report that incorporating fundamental features, such as day-ahead forecasts of renewable generation and load, offers no or very limited improvement \cite{Short, Probab, Forecasting, Understanding, Simulation, 24Multivariate}, thereby motivating that using only the last price as input may \textcolor{black}{yield similar performance to using it together with additional exogenous features.}
However, this assumption does not consider the rich information available in the orderbook. A wide range of orderbook features, such as price percentiles, price momentum, and traded volumes, are not explored and could potentially enhance forecasting performance.

In the context of intraday electricity price forecasting, prior works primarily rely on classical statistical methods,  such as linear regression and its variants \cite{Trading, Probab, Understanding, Econometric}, while more recent studies explore deep learning approaches, such as Multi-Layer Perceptron (MLP), Long Short-Term Memory (LSTM), and Transformer variants \cite{Forecasting, 23Multivariate, Towards, intra, yu2025orderfusion, yu2026deeplearningelectricityprice}, to better capture non-linear patterns in electricity prices. Most of these works have primarily focused on the German market, motivated by its high liquidity and large market size \cite{Trading, 24Multivariate}. Concurrently, there has been a notable shift in research focus from hourly (60-minute) products to quarter-hourly (15-minute) products \cite{Probab, 23Multivariate, 24Multivariate, Simulation}, which provide finer temporal resolution. However, existing studies typically focus on a single type of model, country, and product type, resulting in a fragmented view of model performance. This highlights the need for a unified benchmarking study that systematically compares various machine learning models across countries and product types.

\begin{figure*}[!t]
\begin{center}
    \centering
    \includegraphics[width=1\linewidth]{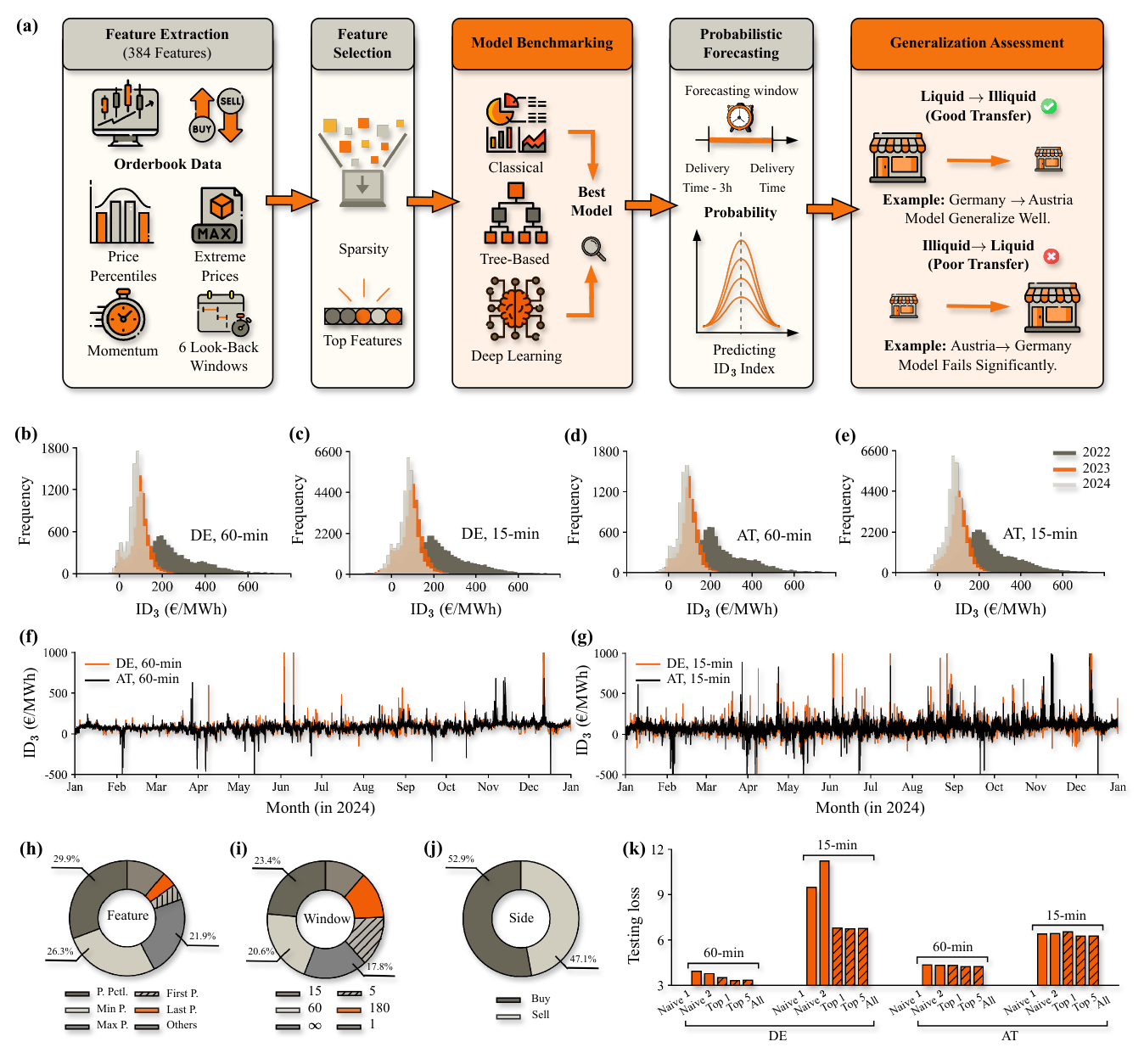}
\caption{
\textcolor{black}{\textbf{(a)} Overview of the workflow: extracting 384 orderbook features, performing sparse feature selection, benchmarking classical, tree-based, and deep models, producing probabilistic forecasts of the ID$_3$ index, and assessing cross-market and cross-product generalization. }
\textbf{(b)}-\textbf{(e)} 
Histograms of 60-min and 15-min ID$_3$ from Germany and Austria.  The price indices exhibit high skewness and dispersion during the energy crisis in 2022, gradually reverting to a more stable distribution in 2023 and 2024. 
\textbf{(f)}-\textbf{(g)} ID$_3$ trajectories in 2024 (range limited to [–500, 1000] €/MWh for better visual comparison). Volatility increases in the order: AT, 60-min $<$ DE, 60-min $<$ AT, 15-min $<$ DE, 15-min.
\textbf{(h)}-\textbf{(j)} Distribution of absolute feature importance by feature type, look-back window size, and market side, respectively.  
\textbf{(k)} Testing loss (AQL) for different feature sets across countries and resolutions.
}
\label{overview*}
\end{center}
\end{figure*}

As most prior studies focus on a single country and product type, it remains unclear whether the selected features and trained models generalize well across different settings. For example, a feature set optimized for the Austrian market may not perform equally well in Germany, and a model trained on 15-min products may fail to capture the dynamics of 60-min prices. This raises important questions about cross-country and cross-product-type generalization. Thus, a systematic investigation into such generalizability is necessary to understand the robustness of derived features, support the transferable development of trained models, and offer actionable insights to stakeholders operating across multiple European markets.

In this paper, we extract 384 features from the orderbook and select the optimal features (Section \ref{Feature_Extraction_Selection}). Then, we provide a comprehensive benchmarking study using classical statistical models, tree-based ensembles, and deep learning models
(Section \ref{model_comparison}). 
Lastly, we assess the cross-country and cross-product-type generalization of the derived optimal features and trained models (Section \ref{Generalization}). We reveal an asymmetric phenomenon: while the optimal feature set and trained model derived from a more liquid market transfer well to a less liquid one, the reverse does not hold. \textcolor{black}{An overview of the workflow is illustrated in Fig.~\ref{overview*} \textbf{(a)}.} Our main contributions are summarized as follows:

\begin{itemize}
    \item We extract an exhaustive set of 384 statistical features from the orderbook, including price percentiles, extreme prices, and VWAPs, and reveal a set of powerful features.

    \item We present a comprehensive benchmark of probabilistic forecasting performance using multiple machine learning models across two countries (Germany and Austria) and two product types (60-min and 15-min).

    \item We systematically assess the generalizability across countries and product types. Our analysis reveals an asymmetric generalization phenomenon.
\end{itemize}

\section{Preliminary}
\label{preliminary}

The forecasting target is the widely used ID$_3$, as visualized in Fig. \ref{overview*} \textcolor{black}{\textbf{(b)-\textbf{(g)}}}.  The \( \mathrm{ID}_3 \) is defined as the VWAP of trades executed within a specific time window before delivery:
\begin{equation}
\mathrm{ID}_3
\;=\;
\frac{\sum\limits_{s \in S} \sum\limits_{t \in \mathcal{T}_{f}} P^{s}_{t} \, V^{s}_{t}}
     {\sum\limits_{s \in S} \sum\limits_{t \in \mathcal{T}_{f}} V^{s}_{t}},
\end{equation}
where the market side \( s \in S = \{+, -\} \) corresponds to buy and sell orders, respectively. 
The forecasting time is defined as \( t_f = t_d - \Delta \), with \( t_d \) denoting the delivery time and \( \Delta = 180\,\mathrm{min} \) representing the lead time specific to \( \mathrm{ID}_3 \).
The transaction time is defined as \( t \in \mathcal{T}_{f} = [t_f,\ t_d - \delta_m] \), where \( \mathcal{T}_{f} \) is the forecasting (trading) window.
\textcolor{black}{$\delta_m$ denotes the market-area–specific cutoff (minutes) set by EPEX that defines the end of the ID$_3$ trading window; equivalently, trades after $t_d-\delta_m$ are excluded from the index (e.g., $\delta_m=0$ for AT, $30$ for DE, and $60$ for DK)}
\footnote{\( \delta_m \) can be retrieved from   \href{https://www.epexspot.com/}{EPEX Spot download center}.}.
Here, \( P^{s}_{t} \) and \( V^{s}_{t} \) denote the price and traded volume, respectively.

\section{Feature Extraction and Selection}
\label{Feature_Extraction_Selection}

\subsection{Feature Extraction}

We extract an exhaustive set of features from both the buy (\(+\)) and sell (\(-\)) sides across multiple look-back windows 
\(
\mathcal{T}_{w} = [t_f - \delta_w, \, t_f]
\), where \( \delta_w \in \{1, 5, 15, 60, 180, \infty\} \) (in minutes), and \( \infty \) denotes the full available trading history. 
The full list of extracted features is summarized in Table~\ref{tab:exhfeatures}. 
If no trades are recorded within a given window (e.g., \( \delta_w = 1 \)), we fall back to the next longer window (e.g., \( \delta_w = 5 \)) to extract features. 
If no trades are observed within the full history window (\( \delta_w = \infty \)), the corresponding sample is discarded.
Feature types include price and volume statistics (e.g., min, max, mean, percentiles), with percentile levels 
\( p \in \mathcal{P} = \{10\%, 25\%, 45\%, 50\%, 55\%, 75\%, 90\%\} \).
\textcolor{black}{For example, \textit{Price Percentile}$\;\big|^{+}_{[t_f-15\,\mathrm{min},\,t_f],\,50\%}$ denotes the median (\(50\%\) quantile) of prices on the buy side ($+$) within the 15-minute look-back window. }

\begin{table}[t]
\caption{Extracted features and definitions.}
\label{tab:exhfeatures}
\begin{center}
\begin{tabular}{ll}
\toprule
\textbf{Feature} & \textbf{Mathematical Definition} \\
\midrule
Price Percentile$\;\big|^{s}_{\mathcal{T}_w,\,p}$ & $\operatorname*{percentile}\limits_{t\in\mathcal{T}_{w}, p} P^{s}_t$ \\
Min Price$\;\big|^{s}_{\mathcal{T}_w}$    & $\min\limits_{t\in\mathcal{T}_w} P^{s}_t$ \\
Max Price$\;\big|^{s}_{\mathcal{T}_w}$    & $\max\limits_{t\in\mathcal{T}_w} P^{s}_t$ \\
First Price$\;\big|^{s}_{\mathcal{T}_w}$  & $\operatorname*{first}\limits_{t\in\mathcal{T}_{w}} P^{s}_t$ \\
Last Price$\;\big|^{s}_{\mathcal{T}_w}$   & $\operatorname*{last}\limits_{t\in\mathcal{T}_{w}} P^{s}_t$ \\
Mean Price$\;\big|^{s}_{\mathcal{T}_w}$   & $\bar P^{s}_{\mathcal{T}_w}$ \\
Price Volatility$\;\big|^{s}_{\mathcal{T}_w}$ & $\sqrt{\dfrac{1}{n^{s}_{\mathcal{T}_w}}\sum\limits_{t\in\mathcal{T}_w}\big(P^{s}_t-\bar P^{s}_{\mathcal{T}_w}\big)^2}$ \\
Delta Price$\;\big|^{s}_{\mathcal{T}_w}$ & $\operatorname*{last}\limits_{t\in\mathcal{T}_{w}} P^{s}_t - \operatorname*{first}\limits_{t\in\mathcal{T}_{w}} P^{s}_t$ \\
Volume Percentile$\;\big|^{s}_{\mathcal{T}_w,\,p}$ & $\operatorname*{percentile}\limits_{t\in\mathcal{T}_{w}, p} V^{s}_t$ \\
Min Volume$\;\big|^{s}_{\mathcal{T}_w}$   & $\min\limits_{t\in\mathcal{T}_w} V^{s}_t$ \\
Max Volume$\;\big|^{s}_{\mathcal{T}_w}$   & $\max\limits_{t\in\mathcal{T}_w} V^{s}_t$ \\
First Volume$\;\big|^{s}_{\mathcal{T}_w}$  & $\operatorname*{first}\limits_{t\in\mathcal{T}_{w}} V^{s}_t$ \\
Last Volume$\;\big|^{s}_{\mathcal{T}_w}$   & $\operatorname*{last}\limits_{t\in\mathcal{T}_{w}} V^{s}_t$ \\
Mean Volume$\;\big|^{s}_{\mathcal{T}_w}$  & $\bar V^{s}_{\mathcal{T}_w}$ \\
Volume Volatility$\;\big|^{s}_{\mathcal{T}_w}$ & $\sqrt{\dfrac{1}{n^{s}_{\mathcal{T}_w}}\sum\limits_{t\in\mathcal{T}_w}\big(V^{s}_t-\bar V^{s}_{\mathcal{T}_w}\big)^2}$ \\
Delta Volume$\;\big|^{s}_{\mathcal{T}_w}$ & $\operatorname*{last}\limits_{t\in\mathcal{T}_{w}} V^{s}_t - \operatorname*{first}\limits_{t\in\mathcal{T}_{w}} V^{s}_t$\\
Sum Volume$\;\big|^{s}_{\mathcal{T}_w}$ & $\sum\limits_{t\in\mathcal{T}_w} V^{s}_t$ \\
Trade Count$\;\big|^{s}_{\mathcal{T}_w}$ & $n^{s}_{\mathcal{T}_w}$ \\
VWAP$\;\big|^{s}_{\mathcal{T}_w}$ & $\dfrac{\sum\limits_{t \in \mathcal{T}_w} P^{s}_t \, V^{s}_t}{\sum\limits_{t \in \mathcal{T}_w} V^{s}_t}$ \\
Momentum$\;\big|^{s}_{\mathcal{T}_w}$ & $\dfrac{\operatorname*{last}\limits_{t\in\mathcal{T}_{w}} P^{s}_t-\mathrm{VWAP}^{s}}{\mathrm{VWAP}^{s}}$ \\
\bottomrule
\end{tabular}
\end{center}
\end{table}

\subsection{Feature Selection}

The extracted feature set may contain redundant or noisy features that harm generalization. 
Following prior works in utilizing \( \ell_1 \)-penalized  linear regression, also known as Least Absolute Shrinkage and Selection Operator (LASSO), to encourage sparse feature sets for pointwise prediction \cite{Understanding}, 
we extend this idea to the probabilistic forecasting setting by applying \( \ell_1 \)-penalized Linear Quantile Regression (LQR) \textcolor{black}{to provide a transparent and interpretable baseline for identifying informative orderbook-derived features}.

Given an input feature matrix \( X_i \in \mathbb{R}^{N \times D} \) and target quantile vector \( y_{i,\tau} \in \mathbb{R}^N \), we estimate the coefficient vector \( \boldsymbol{\beta} \in \mathbb{R}^D \) by solving the following optimization problem:
\begin{equation}
\hat{\boldsymbol{\beta}} = \arg\min_{\boldsymbol{\beta}} \; L_\tau(y_{i,\tau}, X_i \boldsymbol{\beta}) + \alpha \|\boldsymbol{\beta}\|_1
\end{equation}
where \( L_\tau(\cdot) \) denotes the quantile loss:
\begin{equation}
L_\tau(y_{i,\tau}, X_i \boldsymbol{\beta}) = \sum_{n=1}^N \left( y_{i,\tau}^{(n)} - X_i^{(n)} \boldsymbol{\beta} \right) \cdot \left( \tau - \mathbb{I}\{ y_{i,\tau}^{(n)} < X_i^{(n)} \boldsymbol{\beta} \} \right)
\end{equation}
The hyperparameter \( \alpha > 0 \) controls the degree of sparsity by penalizing the absolute magnitudes of the coefficients and is optimized based on validation (quantile) loss. After optimization, only features with non-zero coefficient magnitudes are retained, yielding a reduced sparse feature matrix \( X_i^{(\ell_1)} \in \mathbb{R}^{N \times D_{(\ell_1)}} \), where \( D_{(\ell_1)} \ll D \), that serves as the input to downstream quantile forecasting models.

\section{Model Comparison}
\label{model_comparison}

Based on the optimal feature set identified in the previous step, we compare several machine learning models, spanning classical statistical models, tree-based ensembles, and deep learning models. 

\subsection{Classical Statistical Models} 

\begin{itemize}
    \item \textbf{Linear Quantile Regression (LQR).}  
    LQR models conditional quantiles as linear functions of the input variables. It is highly interpretable and computationally efficient, making it well-suited for high-frequency forecasting tasks~\cite{LQR}. LQR has been widely adopted in the context of intraday electricity price forecasting due to its simplicity and fast training time.

    \item \textbf{Quantile K-Nearest Neighbors (QKNN).}  
    QKNN performs non-parametric quantile regression by computing empirical quantiles from the nearest neighbors in feature space. It makes no assumptions about the underlying data distribution and is capable of capturing strong local nonlinearities, which can be useful in modeling complex market dynamics~\cite{QKNN}.
\end{itemize}

\subsection{Tree-Based Ensemble Learning Models}

\begin{itemize}
    \item \textbf{Quantile LightGBM (QLGBM).}  
    QLGBM extends LightGBM to support quantile regression via gradient boosting and histogram-based tree construction. It efficiently handles large-scale data and captures complex feature interactions~\cite{lgbm}. QLGBM is a widely used model for day-ahead electricity price forecasting due to its speed and robustness across diverse feature sets.

    \item \textbf{Quantile XGBoost (QXGB).}  
    QXGB adapts XGBoost for quantile objectives using regularized decision tree ensembles. It is highly effective at modeling non-linear relationships and capturing long-range dependencies~\cite{xgboost}. XGBoost is also a commonly used model for day-ahead electricity price forecasting.
\end{itemize}

\subsection{Deep Learning Models} 

\begin{itemize}
    \item \textbf{Quantile Multi--layer Perceptron (QMLP).}  
    QMLP applies feedforward neural networks to directly estimate conditional quantiles. It learns complex non-linear mappings and scales well with high-dimensional inputs~\cite{MLP}. QMLP has become a common deep learning baseline for intraday electricity price forecasting.

    \item \textbf{Quantile Kolmogorov--Arnold Networks (QKAN).}  
    QKAN is a recent neural architecture based on the Kolmogorov--Arnold representation theorem, which approximates multivariate functions using compositions of univariate functions. It is designed to learn highly expressive, structured representations~\cite{KAN}. To the best of our knowledge, this work represents the first application of QKAN in intraday electricity price forecasting.
\end{itemize}

\begin{table}[t]
\caption{Hyperparameter search range.}
\label{tab:hyperparams_models}
\begin{center}
\begin{small}
\begin{tabular}{ll}
\toprule
\textbf{Model} & \textbf{Search Range} \\
\midrule

LQR &
\begin{tabular}[t]{@{}l@{}}
\(\ell_1\) regularization: [1e-8, 1]
\end{tabular}
\\
\midrule

QKNN &
\begin{tabular}[t]{@{}l@{}}
n\_neighbors: [5, 100] \\
distance\_metric: \{euclidean, manhattan\} \\
weights: \{uniform, distance\}
\end{tabular}
\\
\midrule

QLGBM &
\begin{tabular}[t]{@{}l@{}}
n\_estimators: [50, 500] \\
max\_depth: [3, 12] \\
learning\_rate: [1e-3, 1e-1] \\
subsample: [0.5, 1.0] \\
colsample\_by\_tree: [0.5, 1.0] \\
reg\_lambda: [0.0, 10.0]
\end{tabular}
\\
\midrule

QXGB &
\begin{tabular}[t]{@{}l@{}}
n\_estimators: [50, 500] \\
max\_depth: [3, 12] \\
learning\_rate: [1e-3, 1e-1] \\
reg\_alpha: [0.0, 5.0] \\
reg\_lambda: [0.0, 10.0]
\end{tabular}
\\
\midrule

QMLP &
\begin{tabular}[t]{@{}l@{}}
hidden\_size: [32, 1024] \\
n\_layers: [2, 6] \\
dropout\_rate: [0.0, 0.5] \\
learning\_rate: [1e-5, 1e-1] \\
batch\_size: [64, 1024]
\end{tabular}
\\
\midrule

QKAN &
\begin{tabular}[t]{@{}l@{}}
kan\_units: [32, 1024] \\
n\_layers: [2, 6] \\
grid\_intervals: [5, 16] \\
spline\_order: [2, 4] \\
learning\_rate: [1e-5, 1e-1] \\
batch\_size: [64, 1024]
\end{tabular}
\\
\bottomrule
\end{tabular}
\end{small}
\end{center}
\end{table}

\subsection{Evaluation Metrics} 
\label{sec:metrics}
Model performance is evaluated using probabilistic and pointwise metrics. For probabilistic forecasting, we employ the Average Quantile Loss (AQL) and the Average Quantile Crossing Rate (AQCR). For pointwise forecasting, we report the Root Mean Squared Error (RMSE), Mean Absolute Error (MAE), and the Coefficient of Determination ($\text{R}^2$).

\begin{itemize}

\item \textbf{Average Quantile Loss (AQL).}  
AQL is employed to jointly evaluate the accuracy of multiple quantiles \cite{yu2025orderfusion}. It aggregates the quantile loss over all quantile levels:
\begin{equation}
    \mathrm{AQL} = \frac{1}{ N|\mathcal{Q}|} \sum_{i=1}^N \sum_{\tau \in \mathcal{Q}}  L_\tau(y_i,  \hat{y}_{i, \tau}),
\end{equation}
where \( y_i \) is the true price, \( \hat{y}_{i, \tau} \) denotes the predicted quantile, and the pinball loss \( L_\tau \) is defined as:
\begin{equation}
    L_\tau(y_i, \hat{y}_{i, \tau}) = 
    \begin{cases} 
      \tau \cdot (y_i -  \hat{y}_{i, \tau}), & \text{if } y_i \geq  \hat{y}_{i, \tau}, \\
      (1 - \tau) \cdot ( \hat{y}_{i, \tau} - y_i), & \text{otherwise}.
    \end{cases}
\end{equation}
The quantile loss penalizes underestimation more heavily at higher quantiles and overestimation more heavily at lower quantiles.

\begin{table*}[t]
\caption{Top 5 features per market, product type, and quantile. 
}
\label{tab:top_feature}
\begin{center}
\begin{tabular}{ccclllll}
\toprule
\textbf{Market} & \textbf{Product Type} & \textbf{Quantile} & & & \textbf{Top 5 features} & & \\
\midrule
DE & 60-min & 0.1 & 
\textcolor{black}{Min P.$\;\big|^{+}_{\mathcal{T}_{15}}$} &
Min P.$\;\big|^{+}_{\mathcal{T}_{60}}$ &
Min P.$\;\big|^{+}_{\mathcal{T}_{\infty}}$ &
Max P.$\;\big|^{-}_{\mathcal{T}_{\infty}}$ &
First P.$\;\big|^{-}_{\mathcal{T}_{1}}$ \\
 &  & 0.5 & \textcolor{black}{P. Pctl.$\;\big|^{-}_{\mathcal{T}_{5},\,90\%}$} &
P. Pctl.$\;\big|^{+}_{\mathcal{T}_{15},\,10\%}$ &
P. Pctl.$\;\big|^{-}_{\mathcal{T}_{15},\,75\%}$ &
Min P.$\;\big|^{+}_{\mathcal{T}_{15}}$ &
P. Pctl.$\;\big|^{+}_{\mathcal{T}_{5},\,10\%}$ \\
 &  & 0.9 & \textcolor{black}{Max P.$\;\big|^{-}_{\mathcal{T}_{15}}$} &
P. Pctl.$\;\big|^{+}_{\mathcal{T}_{\infty},\,10\%}$ &
P. Pctl.$\;\big|^{+}_{\mathcal{T}_{5},\,10\%}$ &
Max P.$\;\big|^{-}_{\mathcal{T}_{60}}$ &
Max P.$\;\big|^{-}_{\mathcal{T}_{180}}$ \\

\cmidrule(lr){2-8}
& 15-min & 0.1 & \textcolor{black}{Min P.$\;\big|^{+}_{\mathcal{T}_{60}}$} &
Min P.$\;\big|^{+}_{\mathcal{T}_{180}}$ &
Max P.$\;\big|^{-}_{\mathcal{T}_{1}}$ &
Mean P.$\;\big|^{-}_{\mathcal{T}_{5}}$ &
Min P.$\;\big|^{+}_{\mathcal{T}_{15}}$ \\
&  & 0.5 & \textcolor{black}{P. Pctl.$\;\big|^{+}_{\mathcal{T}_{60},\,10\%}$} &
P. Pctl.$\;\big|^{-}_{\mathcal{T}_{15},\,90\%}$ &
Min P.$\;\big|^{+}_{\mathcal{T}_{15}}$ &
P. Pctl.$\;\big|^{-}_{\mathcal{T}_{60},\,45\%}$ &
P. Pctl.$\;\big|^{+}_{\mathcal{T}_{60},\,25\%}$ \\
&  & 0.9 & \textcolor{black}{Max P.$\;\big|^{-}_{\mathcal{T}_{60}}$} &
P. Pctl.$\;\big|^{-}_{\mathcal{T}_{60},\,90\%}$ &
Min P.$\;\big|^{+}_{\mathcal{T}_{1}}$ &
Max P.$\;\big|^{-}_{\mathcal{T}_{15}}$ &
Max P.$\;\big|^{-}_{\mathcal{T}_{180}}$ \\
\cmidrule(lr){1-8}
AT & 60-min & 0.1 & \textcolor{black}{Min P.$\;\big|^{+}_{\mathcal{T}_{15}}$} &
Min P.$\;\big|^{+}_{\mathcal{T}_{\infty}}$ &
P. Pctl.$\;\big|^{-}_{\mathcal{T}_{5},\,45\%}$ &
Min P.$\;\big|^{-}_{\mathcal{T}_{5}}$ &
First P.$\;\big|^{-}_{\mathcal{T}_{5}}$ \\
&  & 0.5 & \textcolor{black}{Last P.$\;\big|^{-}_{\mathcal{T}_{\infty}}$} &

Min P.$\;\big|^{+}_{\mathcal{T}_{5}}$ &
P. Pctl.$\;\big|^{-}_{\mathcal{T}_{5},\,45\%}$ &
Last P.$\;\big|^{+}_{\mathcal{T}_{\infty}}$ &
Max P.$\;\big|^{-}_{\mathcal{T}_{1}}$ \\ 
&  & 0.9 & \textcolor{black}{Max P.$\;\big|^{-}_{\mathcal{T}_{180}}$} &
P. Pctl.$\;\big|^{+}_{\mathcal{T}_{60},\,75\%}$ &
Max P.$\;\big|^{-}_{\mathcal{T}_{1}}$ &
First P.$\;\big|^{+}_{\mathcal{T}_{15}}$ &
P. Pctl.$\;\big|^{-}_{\mathcal{T}_{1},\,10\%}$ \\
\cmidrule(lr){2-8}
& 15-min & 0.1 & \textcolor{black}{Min P.$\;\big|^{-}_{\mathcal{T}_{1}}$} &
P. Pctl.$\;\big|^{+}_{\mathcal{T}_{180},\,10\%}$ &
Min P.$\;\big|^{+}_{\mathcal{T}_{15}}$ &
P. Pctl.$\;\big|^{+}_{\mathcal{T}_{\infty},\,10\%}$ &
Min P.$\;\big|^{+}_{\mathcal{T}_{180}}$ \\
& & 0.5 & \textcolor{black}{Mean P.$\;\big|^{+}_{\mathcal{T}_{\infty}}$} &
Last P.$\;\big|^{-}_{\mathcal{T}_{\infty}}$ &
VWAP$\;\big|^{+}_{\mathcal{T}_{1}}$ &
VWAP$\;\big|^{-}_{\mathcal{T}_{180}}$ &
Mean P.$\;\big|^{+}_{\mathcal{T}_{15}}$ \\
&  & 0.9 & \textcolor{black}{Max P.$\;\big|^{+}_{\mathcal{T}_{1}}$} &
Max P.$\;\big|^{+}_{\mathcal{T}_{180}}$ &
Max P.$\;\big|^{-}_{\mathcal{T}_{180}}$ &
Max P.$\;\big|^{-}_{\mathcal{T}_{1}}$ &
Max P.$\;\big|^{+}_{\mathcal{T}_{\infty}}$ \\

\bottomrule
\end{tabular}
\end{center}
\end{table*}

\item \textbf{Average Quantile Crossing Rate (AQCR).}  
AQCR quantifies the frequency of quantile crossing violations \cite{quantilecrossing}, i.e., instances where a lower quantile prediction exceeds a higher quantile prediction. For each sample \( i \), and any quantile pair \( (\tau_l, \tau_u) \) with \( \tau_l < \tau_u \), the crossing indicator is defined as:
\begin{equation}
    C_{\tau_l, \tau_u}(\hat{y}_{l, i}, \hat{y}_{u, i}) = \mathbb{I}(\hat{y}_{l, i} > \hat{y}_{u, i}),
\end{equation}
where \( \mathbb{I}(\cdot) \) is the indicator function. The overall AQCR is then computed as:
\begin{equation}
    \text{AQCR} = \frac{1}{N} \sum_{i=1}^N C_{\tau_l, \tau_u}(\hat{y}_{l, i}, \hat{y}_{u, i}).
\end{equation}
A lower AQCR indicates better consistency of quantile predictions, with fewer violations across quantile levels.

\item \textbf{Root Mean Squared Error (RMSE).}  
RMSE evaluates the overall predictive quality and is sensitive to outliers:
\begin{equation}
    \text{RMSE} = \sqrt{\frac{1}{N} \sum_{i=1}^N (y_i - \hat{y}_i)^2}.
\end{equation}

\item \textbf{Mean Absolute Error (MAE).}  
MAE measures the average magnitude of the prediction errors:
\begin{equation}
    \text{MAE} = \frac{1}{N} \sum_{i=1}^N |y_i - \hat{y}_i|.
\end{equation}

\item \textbf{Coefficient of Determination ($\text{R}^2$).}  
The $\text{R}^2$ score quantifies the proportion of variance in the target variable explained by the model:
\begin{equation}
    \text{R}^2 = 1 - \frac{\sum_{i=1}^N (y_i - \hat{y}_i)^2}{\sum_{i=1}^N (y_i - \bar{y})^2},
\end{equation}
where \( \bar{y} \) is the mean of the true values.

\end{itemize}

\begin{table*}[ht]
\caption{Performance comparison of machine learning models. }
\label{tab:baselines}
\begin{center}
\begin{tabular}{cclcccccc}
\toprule
\textbf{Market} & \textbf{Product Type} & \textbf{Model} & \textbf{AQL} $\downarrow$& \textbf{AQCR} $\downarrow$& \textbf{RMSE} $\downarrow$& \textbf{MAE} $\downarrow$& \textbf{R2} $\uparrow$\\
\midrule
DE & 60-min & LQR$^{\dagger}$   & 3.42±0.00 & 0.05±0.00 & 27.94±0.00 & 10.05±0.00 & 0.87±0.00 \\
   &        & QKNN$^{\dagger}$  & 3.51±0.00 & 0.06±0.00 & 28.41±0.00 & 10.33±0.00 & 0.86±0.00 \\
   &        & QLGBM & 3.62±0.00 & 0.14±0.03 & 29.88±0.14 & 11.45±0.29 & 0.85±0.01  \\
   &        & QXGB  & 3.59±0.00 & 0.08±0.03 & 28.75±0.36 & 10.95±0.10 & 0.86±0.01  \\
   &        & QMLP  & \textbf{3.30±0.01} & \textbf{0.01±0.00} & \textbf{27.67±0.24} & \textbf{10.02±0.02} & \underline{0.87±0.00} \\
   &        & QKAN  & \underline{3.32±0.02} & \underline{0.01±0.00} & \underline{27.69±0.17} & \underline{10.06±0.05} & \textbf{0.87±0.00} \\

\cmidrule(lr){2-8}
   & 15-min & LQR$^{\dagger}$   & 6.79±0.00 & 0.09±0.00 & 52.68±0.00 & 18.41±0.00 & 0.69±0.00 \\
   &        & QKNN$^{\dagger}$  & 6.80±0.00 & 0.10±0.00 & 52.78±0.00 & 18.56±0.00 & 0.69±0.00 \\
   &        & QLGBM & 6.82±0.05 & 0.19±0.09 & 52.93±0.37 & 18.37±0.20 & 0.69±0.01 &  \\
   &        & QXGB  & 6.75±0.04 & 0.14±0.05 & 52.40±0.41 & 17.97±0.33 & 0.70±0.01 &  \\

   &        & QMLP  & \underline{6.56±0.03} & \textbf{0.01±0.00} & \textbf{50.40±0.32} & \textbf{17.84±0.28} & \textbf{0.72±0.00} \\
   &        & QKAN  & \textbf{6.56±0.02} & \underline{0.01±0.00} & \underline{50.42±0.27} & \underline{17.88±0.09} & \underline{0.72±0.00} \\

\cmidrule(lr){1-8}
AT & 60-min & LQR$^{\dagger}$   & 4.33±0.00 & 0.05±0.00 & 28.61±0.00 & 11.91±0.00 & 0.80±0.00 \\
   &        & QKNN$^{\dagger}$  & 4.46±0.00 & 0.04±0.00 & 29.29±0.00 & 12.23±0.00 & 0.79±0.00 \\
   &        & QLGBM & 4.49±0.05 & 0.10±0.04 & 29.41±0.42 & 12.14±0.17 &  0.78±0.01 \\
   &        & QXGB  & 4.45±0.03 & 0.04±0.01 & 28.90±0.38 & 11.97±0.10 &  0.79±0.01 \\
   &        & QMLP  & \underline{4.20±0.01} & \textbf{0.01±0.00} & \underline{28.33±0.23} & \textbf{11.69±0.06} & \textbf{0.80±0.00} \\
   &        & QKAN  & \textbf{4.19±0.01} & \underline{0.01±0.00} & \textbf{28.27±0.14} & \underline{11.75±0.10} & \underline{0.80±0.00} \\
\cmidrule(lr){2-8}

   & 15-min & LQR$^{\dagger}$   & 6.44±0.00 & 0.04±0.00 & 52.99±0.00 & 17.82±0.00 & 0.57±0.00 \\
   &        & QKNN$^{\dagger}$  & 6.44±0.00 & 0.09±0.00 & 52.02±0.00 & 17.84±0.00 & 0.57±0.00 \\
   &        & QLGBM & 6.38±0.11 & 0.05±0.01 & 51.97±0.24 & 17.42±0.18 & 0.57±0.01 \\
   &        & QXGB  & 6.47±0.08 & 0.07±0.02 & 51.67±0.18 & 17.32±0.15 & 0.58±0.01 \\
   &        & QMLP  & \textbf{6.22±0.06} & \textbf{0.01±0.00} & \underline{51.24±0.22} & \underline{17.18±0.21} & \textbf{0.58±0.00} \\
   &        & QKAN  & \underline{6.22±0.08} & \underline{0.01±0.00} & \textbf{51.15±0.11} & \textbf{17.09±0.14} & \underline{0.58±0.00} \\
\bottomrule
\end{tabular}
\end{center}
\end{table*}

\section{Generalization Assessment}
\label{Generalization}

To assess the model's generalizability  across countries and product types, we conduct two sets of experiments: (1) \textit{cross-country generalization}, where the model is transferred between DE and AT; and (2) \textit{cross-product-type generalization}, where the model is transferred between 60-min and 15-min products.
For both experiments, the following three transfer strategies are applied:
\begin{itemize}
    \item \textbf{A $\rightarrow$ A:} Use the optimal feature set derived from domain A; train and optimize the model on data from domain A; test it on domain A.
    
    \item \textbf{B $\rightarrow$ A:} Use the optimal feature set derived from domain B; train and optimize the model on data from domain B; test on domain A.
    
    \item \textbf{A} + \textbf{B $\rightarrow$ A:} Use the union of optimal feature sets from both domains A and B; train and optimize the model on combined data from A and B; test on domain A.
\end{itemize}

Furthermore, we introduce two measures to quantify the phenomenon of \textit{asymmetric generalization}: the \emph{loss ratio} \(\mathcal{L}\) and the \emph{trade-count ratio} \(\mathcal{C}\), defined as follows:
\begin{equation}
\mathcal{L} \;=\;
\frac{\mathrm{AQL}(\text{B}\!\to\!\text{A})}
     {\mathrm{AQL}(\text{A}\!\to\!\text{A})},
\label{eq:lossratio}
\end{equation}
where \(\mathrm{AQL}(\text{B}\!\to\!\text{A})\) and \(\mathrm{AQL}(\text{A}\!\to\!\text{A})\)  are testing loss and can be retrieved from Table~\ref{tab:cross_domain}. A higher value of \(\mathcal{L}\) indicates that a model transferred from domain B performs worse on domain A than a model trained directly on domain A;
\begin{equation}
\mathcal{C} \;=\; \frac{N_{\text{B}}}{N_{\text{A}}},
\label{eq:traderatio}
\end{equation}
where \(N_{\text{A}}\) and \(N_{\text{B}}\) represent the average count of matched trades across testing samples for domains A and B, respectively. These values reflect the market liquidity and are illustrated in Fig.~\ref{fig:transfer_liquidity} \textbf{(a)}, where DE 60-min exhibits the highest trade count (most liquid), and AT 15-min the lowest (least liquid). A higher value of \(\mathcal{C}\) indicates that transfer learning is performed from a more liquid domain to a less liquid one.

\subsection{Cross-Country Generalization}

In this setting, domains A and B refer to countries. Specifically, A can be either DE or AT, and B is the other. We evaluate model generalization across countries for the 60-min and 15-min product types, respectively.

\subsection{Cross-Product-Type Generalization}

In this setting, domains A and B refer to product types. Specifically, A can be either the 60-min or 15-min product, and B is the other. We evaluate model generalization across product types for the DE and AT markets, respectively.

\section{Experiment}

The orderbook is split into training (2022-01-01 to 2024-01-01), validation (2024-01-01 to 2024-07-01), and testing (2024-07-01 to 2025-01-01) periods. The testing window is chosen to examine the model's performance on  more recent, up-to-date data. 
For the 60-min and 15-min products, a prediction is generated every 60 minutes and 15 minutes, respectively. 
\textcolor{black}{Each model is optimized with 100 trials using Optuna, which applies Bayesian optimization for efficient hyperparameter tuning \cite{Optuna}. The corresponding hyperparameter search ranges are summarized in Table~\ref{tab:hyperparams_models}. }

\subsection{Feature Extraction and Selection}
We retain features with non-zero coefficients and analyze their importance by summing the absolute coefficient magnitudes across feature type, look-back window size, and market side, respectively. As shown in Fig.~\ref{overview*} \textbf{(h)}, the most important feature types are \textit{price percentiles} (29.9\%), \textit{minimum prices} (26.3\%), and \textit{maximum prices} (21.9\%), while the volume-based features do not contribute much in terms of coefficient magnitude. 
In Fig.~\ref{overview*} \textbf{(i)}, features extracted from the last 15  and 60 minutes contribute the most (23.4\% and 20.6\%). Surprisingly, the last 1-minute window contributes only 11.3\% of total importance. This may be due to the volatility and noise in short-term trading activity. Fig.~\ref{overview*} \textbf{(j)} shows that buy-side features slightly dominate sell-side features, although the difference is marginal. \textcolor{black}{We note that this feature-importance analysis reflects only linear associations between features and price labels, non-linear models may exploit complex interactions and non-linear transformations, which can shift the relative importance across feature groups. }

To evaluate the impact of feature selection on performance, we rank all features by their absolute coefficient values per market, product type, and quantile. The LQR models are retrained using: only the top 1 feature; the top 5 features; and all the selected features. Additionally, we include two previously reported strong predictors as benchmarks: VWAP over the last 15 minutes (Naive$^1$) and the last price (Naive$^2$). As shown in Fig.~\ref{overview*} \textbf{(k)}, the selected full feature set significantly outperforms both naive baselines, achieving on average {10.53\%} and {11.87\%} lower testing loss compared to Naive$^1$ and Naive$^2$, respectively. Moreover, the top 5 features are often sufficient to match the performance of the full set, indicating redundancy among weaker features. While the top 1 feature yields comparable performance to the top 5 in the 15-minute product in DE, it performs worse in other settings. Therefore, we proceed by using the top 5 features, revealed in Table~\ref{tab:top_feature}, for the downstream forecasting task.

\begin{table*}[ht]
\caption{Cross-domain generalization performance.}
\label{tab:cross_domain}
\begin{center}
\begin{tabular}{ccccccc}

\multicolumn{7}{l}{{Cross-Country Generalization}} \\
\toprule

\textbf{Product Type} & \textbf{Trans. Strategy} & \textbf{AQL} $\downarrow$ & \textbf{AQCR} $\downarrow$& \textbf{RMSE} $\downarrow$& \textbf{MAE} $\downarrow$& \textbf{R2} $\uparrow$ \\

\midrule
60-min & DE $\rightarrow$ DE & \textbf{3.30±0.01} & \textbf{0.01±0.00} & \textbf{27.67±0.24} & \textbf{10.02±0.01} & \textbf{0.87±0.00} \\

& {AT $\rightarrow$ DE} & \cellcolor{orange}{23.84±13.34} & {0.41±0.10} & {35.60±9.64} & {11.60±1.61} & {0.78±0.12} \\

& DE $+$ AT $\rightarrow$ DE & 3.47±0.02 & 0.01±0.00 & 27.91±0.09 & 10.12±0.04 & 0.86±0.01 \\

\cmidrule(lr){2-7}

 & AT $\rightarrow$ AT & 4.20±0.01 & 0.01±0.00 & 28.33±0.23 & 11.69±0.06 & 0.80±0.00 \\
& {DE $\rightarrow$ AT} & \cellcolor{custombeige}{4.20±0.01} & {0.02±0.00} & {28.85±0.06} & {11.98±0.04} & {0.80±0.00} \\
& DE $+$ AT $\rightarrow$ AT & \textbf{4.15±0.01} & \textbf{0.01±0.00} & \textbf{28.23±0.05} & \textbf{11.55±0.07} & \textbf{0.80±0.00} \\
\midrule

15-min & DE $\rightarrow$ DE & \textbf{6.56±0.03} & \textbf{0.01±0.00} & \textbf{50.40±0.32} & \textbf{17.84±0.28} & \textbf{0.72±0.00} \\

& {AT $\rightarrow$ DE} & \cellcolor{orange}{80.15±11.23} & {0.08±0.01} & {88.99±8.16} & {39.48±2.84} & {0.12±0.16} \\
& DE $+$ AT $\rightarrow$ DE & 8.89±0.41 & 0.06±0.04 & 55.29±2.08 & 21.94±1.33 & 0.66±0.03 \\

\cmidrule(lr){2-7}
 & AT $\rightarrow$ AT & 6.22±0.06 & 0.01±0.00 & 51.24±0.22 & 17.18±0.21 & 0.58±0.00 \\

& {DE $\rightarrow$ AT} & \cellcolor{custombeige}{6.27±0.12} & {0.02±0.04} & {52.89±0.13} & {17.79±0.19} & {0.58±0.00} \\
& DE $+$ AT $\rightarrow$ AT & \textbf{6.20±0.03} & \textbf{0.00±0.00} & \textbf{51.00±0.04} & \textbf{17.04±0.09} & \textbf{0.58±0.00} \\

\multicolumn{7}{l}{} \\

\multicolumn{7}{l}{{Cross-Product-Type Generalization}} \\
\toprule
\textbf{Market} & \textbf{Trans. Strategy} & \textbf{AQL} $\downarrow$& \textbf{AQCR} $\downarrow$& \textbf{RMSE} $\downarrow$& \textbf{MAE} $\downarrow$& \textbf{R2} $\uparrow$\\

\midrule
DE & 60-min $\rightarrow$ 60-min & \textbf{3.30±0.01} & \textbf{0.01±0.00} & \textbf{27.67±0.24} & \textbf{10.02±0.01} & \textbf{0.87±0.00} \\

& {15-min $\rightarrow$ 60-min} & \cellcolor{orange}{4.45±0.42} & {0.04±0.03} & {28.90±0.07} & {10.34±0.11} & {0.86±0.00} \\

& 60-min $+$ 15-min $\rightarrow$ 60-min & 3.57±0.01 & 0.02±0.00 & 27.98±0.12 & 10.13±0.06 & 0.87±0.00 \\

\cmidrule(lr){2-7}

 & 15-min $\rightarrow$ 15-min & 6.56±0.03 & 0.01±0.00 & 50.40±0.32 & 17.84±0.28 & 0.72±0.00 \\
 
& {60-min  $\rightarrow$ 15-min} & \cellcolor{custombeige}{6.59±0.28} & {0.02±0.00} & {50.83±2.39} & {17.89±0.35} & {0.72±0.00} \\

& 60-min $+$ 15-min $\rightarrow$ 15-min & \textbf{6.33±0.02} & \textbf{0.01±0.00} & \textbf{50.10±0.36} & \textbf{17.63±0.07} & \textbf{0.72±0.00} \\

\midrule

AT & 60-min $\rightarrow$ 60-min & \textbf{4.20±0.01} & \textbf{0.01±0.00} & \textbf{28.33±0.23} & \textbf{11.69±0.06} & \textbf{0.80±0.00} \\

& {15-min $\rightarrow$ 60-min} & \cellcolor{orange}{8.17±3.90} & {0.13±0.23} & {29.99±0.24} & {12.41±0.10} & {0.78±0.00} \\

& 60-min $+$ 15-min $\rightarrow$ 60-min & 4.46±0.04 & 0.02±0.00 & 29.47±0.03 & 12.16±0.04 & 0.79±0.00 \\

\cmidrule(lr){2-7}
 & 15-min $\rightarrow$ 15-min & 6.22±0.06 & 0.01±0.00 & 51.24±0.22 & 17.18±0.21 & 0.58±0.00 \\

& {60-min  $\rightarrow$ 15-min} & \cellcolor{custombeige}{6.22±0.09} & {0.01±0.00} & {52.18±0.25} & {17.59±0.11} & {0.58±0.00} \\

& 60-min $+$ 15-min $\rightarrow$ 15-min & \textbf{6.21±0.05} & \textbf{0.00±0.00} & \textbf{51.15±0.05} & \textbf{17.05±0.05} & \textbf{0.58±0.00} \\

\bottomrule
\end{tabular}
\end{center}
\end{table*}

\subsection{Model Comparison} 
The results of the model comparison are illustrated in Table \ref{tab:baselines}, where all metrics are reported as mean±standard deviation over 5 independent runs. 
The best results are in \textbf{bold}. Models marked with $^{\dagger}$ lack random-seed control; thus, the standard deviation is zero.
The units of AQL, RMSE, and MAE are expressed in~\euro{}/\text{MWh}, and AQCR in~\%. We observe substantial variation in AQL across probabilistic forecasting scenarios. Specifically, 
the difficulty in probabilistic forecasting is in the order: DE, 60-min $<$ AT, 60-min $<$ AT, 15-min $<$ DE, 15-min, as indicated by the average AQL across models. 
This order contradicts the volatility order: AT, 60-min $<$ DE, 60-min $<$ AT, 15-min $<$ DE, 15-min, as observed from Fig. \ref{overview*} \textbf{(f)}-\textbf{(g)}.
One possible explanation is that the higher liquidity in the DE market provides richer and more stable predictive features for 60-min products, partially offsetting the impact of volatility. In contrast, the AT 60-min product combines lower volatility with lower liquidity compared to the DE 60-min product, and the lower liquidity increases the difficulty of probabilistic forecasting.

Among the six models compared, the deep learning approaches consistently outperform classical statistical models and tree-based ensembles. In particular, QMLP achieves on average {3.45\%} and {4.59\%} lower AQL than LQR and QKNN, respectively, when averaged across markets and product types. 
Furthermore, QLGBM and QXGB result in {5.08\%} and {4.83\%} higher AQL on average compared to  QMLP. 
In addition, the classical statistical methods and tree-based ensembles exhibit higher AQCR values, ranging from 0.04\% to 0.19\%, indicating more unreliable probabilistic forecasting. This issue is expected to be further magnified when predicting additional quantiles.
We also note that QMLP and QKAN perform nearly identically across all metrics. 
However, QKAN requires approximately 9.7 times longer training time per epoch due to its neural decomposition and multivariate integration structure. Therefore, QMLP offers the best trade-off between computational efficiency and predictive performance and is selected for the downstream generalization assessment.

\subsection{Generalization Assessment}

In the \textit{cross-country experiments}, models trained on the DE market generalize well to the AT market, while the reverse direction results in substantial degradation of performance, as observed from Table \ref{tab:cross_domain}. 
In both 60-min and 15-min settings, separate training achieves the best performance across all metrics when predicting DE prices, while joint training leads to the best or equivalent performance when predicting AT prices. 
Notably, when directly transferring a model trained on AT orderbook data, the AQL increases drastically from 3.30 to 23.84 for the 60-min product and from 6.56 to 80.15 for the 15-min product (highlighted in orange in the table). 
In contrast, DE-trained models maintain similar performance when applied directly to the AT market  (highlighted in gray in the table). 
\textcolor{black}{These results highlight a clear asymmetric phenomenon, consistent with the higher liquidity of the DE market being associated with better transfer toward the less liquid AT market.}

In the \textit{cross-product-type experiments}, models trained on the 60-min product generalize well to the 15-min product, while the reverse direction again leads to inferior performance, as observed from Table \ref{tab:cross_domain}. 
In both DE and AT markets, separate training yields the best results across all metrics when predicting 60-min prices, whereas joint training improves or maintains performance when predicting 15-min prices. 
Notably, directly transferring a model trained on 15-min data results in AQL increases from 3.30 to 4.45 in DE and from 4.20 to 8.17 in AT (highlighted in orange in the table). 
Meanwhile, transferring from 60-min to 15-min retains similar performance compared to separate training (highlighted in gray in the table). 
\textcolor{black}{These results again suggest that the asymmetric phenomenon is associated with liquidity differences, as 60-min products contain more trades.}
In contrast, 15-min products are sparser and more volatile, limiting their generalizability to coarser timescales.

\begin{figure}[!ht]
\begin{center}
    \centering
    \includegraphics[width=1\linewidth]{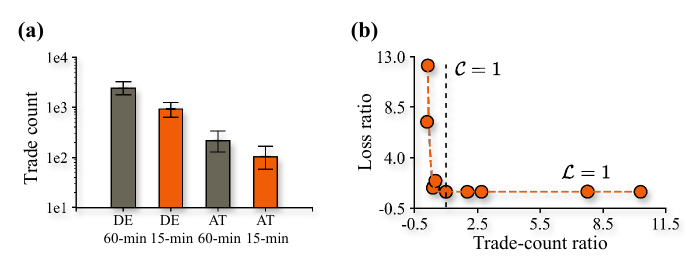}
\caption{
Analysis of model performance against market liquidity.  
\textbf{(a)} Comparison of market liquidity.
\textbf{(b)} Loss ratio versus trade-count ratio.
}
\label{fig:transfer_liquidity}
\end{center}
\end{figure}

Fig.~\ref{fig:transfer_liquidity} \textbf{(b)} shows the scatter of \((\mathcal{C}, \mathcal{L})\) and its empirical fitting curve. For \(\mathcal{C} \ge 1\), where transfer learning is performed from a more liquid domain to a less liquid one, the loss ratio remains at \(\mathcal{L} = 1\), indicating performance equivalent to training directly on the target domain. In contrast, for \(\mathcal{C} < 1\), where transfer occurs from a less liquid domain to a more liquid one, a clear exponential trend is observed: as \(\mathcal{C}\) decreases, the loss ratio \(\mathcal{L}\) increases sharply, indicating worse performance compared to target-only training. These observations confirm the role of liquidity in transfer performance and support the emergence of the asymmetric generalization phenomenon.

\section{Conclusion}
\label{Conclusion}

In this paper, we developed a comprehensive feature set consisting of 384 orderbook features and revealed that price percentiles and extreme prices outperform the previously reported powerful features such as VWAP and last price. Moreover, through model comparison, we find that deep learning models consistently outperform classical statistical models and tree-based ensembles. In particular, QMLPs emerge as a strong baseline for probabilistic forecasting when using domain features. Finally, our generalization assessment uncovers a pronounced asymmetry in transferability: models trained on more liquid markets or products generalize well to less liquid domains, while the reverse transfer leads to substantial performance degradation. These findings underscore the importance of market liquidity in designing better models for probabilistic intraday electricity price forecasting.

\section{Limitation and Future Work}
\label{Limitation_and_future_work}
First, the extracted features in this study are empirical and may benefit from exploring a broader feature set in future work.  Second, as markets become more efficient, simpler indicators such as the last price may become sufficient. We will monitor such developments, particularly as electricity markets transition to full quarter-hourly resolution. 
Third, the hyperparameters are tuned empirically. Additional hyperparameter tuning and a larger number of trials may further improve performance, potentially enabling tree-based models to match the performance of deep learning models. \textcolor{black}{Fourth, we use trade-count as a simple liquidity proxy; examining additional liquidity measures (e.g., bid-ask spread and orderbook depth) and their relationship to asymmetric generalization is an interesting direction for future work.}
\textcolor{black}{Fifth, the features are selected via LQR, which capture only linear effects; extending feature selection to non-linear models is left for future work.}
Lastly, this work focuses on the central regions in Europe; extending the analysis to Nordic markets is worth exploring.

\printbibliography[heading=bibintoc,title=References]

\end{document}